\documentclass[natbib=false, sigconf, authorversion, nonacm]{acmart}
\pdfoutput=1
\usepackage{acronym}
\usepackage{stfloats}

\RequirePackage[datamodel=acmdatamodel, style=acmnumeric, sortcites=true, backend=biber]{biblatex}
\usepackage{etoolbox}

\AtBeginDocument{%
  }

\copyrightyear{2025}
\acmYear{2025}
\begin{document}

\title{Pervasive Sensing for Livestock Health and Activity Monitoring: Current Methods and Techniques}


\author{Jeffrey D Shulkin}
\orcid{0009-0008-5286-7454}
\email{shulkinj@umich.edu}
\affiliation{
	\institution{University of Michigan}
	\city{Ann Arbor}
	\state{Michigan}
	\country{USA}
}

\author{Abhipol Vibhatasilpin}
\orcid{0009-0005-2587-4163}
\email{abhipol@umich.edu}
\affiliation{
	\institution{University of Michigan}
	\city{Ann Arbor}
	\state{Michigan}
	\country{USA}
}

\author{Vedant Adhana}
\orcid{}
\email{atdhana@umich.edu}
\affiliation{
	\institution{University of Michigan}
	\city{Ann Arbor}
	\state{Michigan}
	\country{USA}
}

\renewcommand{\shortauthors}{Shulkin et al.}

\begin{abstract}
Pervasive sensing is transforming health and activity monitoring by enabling continuous and automated data collection through advanced sensing modalities. While extensive research has been conducted on human subjects, its application in livestock remains underexplored. In large-scale agriculture, real-time monitoring of biological signals and behavioral patterns can facilitate early disease detection, optimize feeding and breeding strategies, and ensure compliance with welfare standards. This survey examines key sensing technologies—including structural vibration, radio frequency (RF), computer vision, and wearables—highlighting their benefits and challenges in livestock monitoring. By comparing these approaches, we provide insights into their effectiveness, limitations, and potential for integration into modern smart farming systems. Finally, we discuss research gaps and future directions to advance pervasive sensing in livestock health and activity monitoring.
\end{abstract}





\maketitle
\section{Introduction}
\label{sec::intro}

Smart monitoring of livestock is essential for improving animal health and welfare by detecting early signs of illness or distress and enabling timely interventions. It enhances efficiency and productivity by optimizing feeding schedules, breeding routines, and farm management while reducing costs.

Pervasive health and activity monitoring is emerging as a transformative field at the intersection of ubiquitous computing and healthcare. Recent developments in sensing technologies have enabled computing systems to construct a more complete and insightful analysis of the user's health and provide better guidelines for intervention. 
This domain of study incorporates various cutting-edge techniques such as computer vision, RF reflectometry, structural vibrations, wearables, and smart environments. 
To facilitate continuous extraction of various biological signals, and real-time monitoring of the environment and its occupants.

While much of the exploratory work have been done in humans, the application of pervasive health and activity monitoring techniques is largely unexplored in animals, especially livestocks in largescale agriculture. With real-time stream of biological and activity data, these insights can help facilitate early detection of health issues, optimize feeding and breeding routines, and ensure welfare compliance, ultimately leading to a more efficient and sustainable farming operation. In this survey, we compare and contrast various methods for monitoring the health and activites of livestock in modern smart farms. We specifically focus on the benefits and challenges behind structural vibration, radio frequency (RF), computer vision, and wearable modalities.

The rest of the article is organized as follows. In Section 2, we analyze vibration based methods in both human and livestock monitoring. In Section 3, we document how computer vision methods can be utilized to monitor livestock in a scalable manner. In Section 4, we document the ways in which \texttt{mmWave} and other radio-frequency approaches have been applied to livestock and human health monitoring. Section 5 presents our survey on wearable methods, both for human health monitoring and for tracking the activities of wild animals and livestock. In Section 6, we discuss critical research challenges and propose future research directions. In Section 7, we make some concluding remarks and point out the implications of this research.

\section{Vibration}
\label{sec::vib}

While traditionally used for sensing the health of bridges~\cite{liu.etal2023TelecomDAS, liu.etal2020DamageSensitiveBridgeHealthMonitor, liu.etal2023HierMUD, liu.etal2020DiagnosisAlgorithms}, buildings, and other infrastructure, as well as seismic events, vibration sensors have gained traction as a viable modality for occupancy detection, activity monitoring, and health sensing. In this section, we will explore the various ways in which vibration sensors have been used to capture and analyze different signals on both human subjects and on livestock. 

\subsection{Vibration Sensing in Human Health Monitoring}
\label{subsec::human_vib}

In recent times, vibration sensing has also emerged as a promising approach for monitoring health in humans. By capturing subtle mechanical vibrations transmitted through floors or other surfaces, these sensors offer non-invasive information on a person's physical state and activities whilst preserving user privacy by recording only time-series data lacking directly identifiable features.

Given their ability to detect small and large amplitude events, vibration sensors have been used not only in classifying individual occupancy of a particular space~\cite{pan.etal2014BOESBuilding}, but also in detecting the activity of large crowds during sporting events~\cite{GameVibes, chang.etal2024PosterAbstracta}. Unlike computer vision-based methods, which can potentially generate terabytes of information, relatively sparse deployments of vibration sensors can be used to monitor entire stadiums, placing a much lower data burden on researchers during post-deployment analysis.

In human subjects, structural vibration systems have been used not only to detect whether an individual is occupying a specific space~\cite{pan.etal2014BOESBuilding, asod, zhang.etal2024PosterAbstract, dong.etal2021SocialDistancing}, but also to identify the specific individual~\cite{dong.etal2023StrangerDetection, pan.etal2017FootprintIDIndoor} and classify their activities within that space~\cite{zhang.etal2018VibrationbasedOccupant, oac,bonde2020NoisetolerantContextaware, zhang.etal2018OccupantinducedOffice, pan.etal2018CharacterizingHumana, pan.etal2019FinegrainedRecognition, bonde.etal2019DeskbuddyOffice, fagert.etal2017MonitoringHandwashing}.

An application of this technology is in fall detection and prevention ~\cite{zigel.etal2009VibrationFallDetection, fagert.etal2017CharacterizingLeftright, zhang.etal2024PosterAbstract}. Early studies demonstrate that this can be used to confirm a fall event and also alert caregivers promptly, potentially reducing intervention time in critical situations without recording potentially private information.

Vibration sensing has also been used to perform human gait analysis~\cite{dong.etal2024AmbientFloora, dong.etal2022GaitVibeEnhancing, dong.noh2024StructureagnosticGait, dong.noh2024UbiquitousGaita, dong.etal2024InhomeGaita, dong.etal2023DetectingGait, dong.etal2024RobustPersonalized, mirshekari.etal2018HumanGait, fagert.etal2019VibrationSource, fagert.etal2017CharacterizingLeftright}. Continuous monitoring of the vibration patterns produced by footsteps enables researchers to extract parameters such as step frequency, stride length, and walking stability. Analyzing variations in these patterns over time can help in early diagnosis and health assessment and identify several underlying neurological disorders or musculoskeletal problems ~\cite{dong.etal2024AmbientFloora, dong.noh2024UbiquitousGaita}.

Furthermore, vibration sensors have been utilized to indirectly monitor vital signs~\cite{FreePulse, jia2017monitoring, codling.etal2024FloHRUbiquitous, VALERO2021103037, bonde.etal2017HeartSoleb, bonde.etal2018SeatVibrationa, jia.etal2016HBphoneBedmounteda}. In certain setups, the minute body vibrations associated with heartbeat and respiration can be distinguished from other movement-related signals. This approach minimizes the need for wearable devices and allows for continuous monitoring in everyday environments, making it suitable for elderly or mobility-impaired individuals.

\subsection{Vibration Sensing in Smart Farms}
\label{subsec::smart_farm_vib}

More recently, researchers have investigated the use of structural vibration in order to monitor both the activities and vital signs of animals raised in smart farms. 

Vibration sensing has been specifically been used to monitor the activities and vital signs of sows and piglets during their farrowing period~\cite{pignet, MassHog, PigV2, pigsense, bonde.etal2018StructuralVibration}. In these smart phones, each sow and piglet litter is housed within a tight pen. Given that pigs can weigh between 140 kg and 300 kg, each movement they make exerts pressure on the surrounding pen. By attaching vibration sensors around the pig pen, researchers were able to use these pressure waves to identify when the sows were eating, drinking, and whether the pigs were standing, sitting, or lying on the pen's grate. In addition, unlike vision-based methods, vibration sensing is a privacy-preserving modality, as the time-series signal does not capture personally identifiable information. 

\subsection{Challenges in Health Monitoring via Structural Vibration}
\label{subsec::vib_challenges}

While structural vibration preserves the user's privacy, sensing vital signs and activities purely through vibration presents its own set of challenges. 

First, these vibration waves often propagate through dispersive media, which warps the signals as it travels from the body through the structure. While vibration signal warping is well studied with regards to civil structures, the effects of propagation through man-made structures, such as wood or concrete, is not well known with regards to the often variable signals involved in occupancy detection and health sensing~\cite{codling.etal2024FloHRUbiquitous, codling.etal2023DemoAbstract, SeatBeats, pan.etal2014BOESBuilding}. As a result, systems employing vibration as their primary modality often have to feed their data through machine learning models to achieve reasonable accuracy, which often require large amounts of labeled data for robust training~\cite{pignet, PigV2, pigsense, MassHog, dong.etal2024ContextawareCrowda, GameVibes, dong.noh2024UbiquitousGaita,  chang.etal2024PosterAbstracta, mirshekari.etal2019PhysicsguidedModel}. 

Second, different biometric signals buried within the vibration signal often require vastly different gain settings~\cite{MassHog, pignet, PigV2, codling.etal2024FloHRUbiquitous}. For example, because heart beat vibrations have a very small amplitude, the sensor node extracting this signal must have a high signal gain. However, due to the node's high sensitivity, the heart beat signal can be masked by higher amplitude vibrations, such as nearby footsteps. As a result, the vibration environment must be finely tuned in order to reliably extract the heart-rate signal, which can limit the scenarios in which vibration sensor deployment can succeed.

Third, vibration signals vary not ony based on the structures they propagate through, but also based on the person from whom the signals originate. For example, even given the same structural environment, two individuals can have vastly different heartbeat vibration signals, due in part to the differences in body structure, weight, and other aspects of their health. As a result, training machine learning models for vibration sensing is difficult, as, unlike audio and other modalities, labeled, open source vibration datasets are rare. Different methods have worked to address this gap via pre-training vibration-based models with data from other wave-based modalities, such as audio~\cite{chang.etal2024PosterAbstracta}, but this lack of large open source datasets still means that the accuracy of these models is limited.  

Lastly, sensing both activity and vital signs through the same vibration sensor node is quite difficult~\cite{pigsense}. While previous works have focused on sensing coarse activities or fine-grain vital signs, no work to our knowledge has attempted to sense both at the same time. This is a major gap in smart farm health monitoring, as sensing both a livestock animal's vital signs and specific activity can grant deeper insight into their long term health, as well as provide information on how the specific activity being performed shapes the vital sign signal itself. 

\section{Computer Vision}
\label{sec::cv}

Computer vision (CV) has emerged as a powerful technique for pervasive sensing, enabling persistent and contactless monitoring of activities and behaviors of multiple users in the field of view. Over the years, advancements in convolutional neural networks (CNNs), vision transformer (ViTs), and self-supervised learning have greatly enhanced the accuracy of vision sensing systems, and improved their robustness against real-world uncertainties such as variations in ambient lighting and highly dynamic scene compositions. With camera hardware becoming increasingly prevalent, we examine the current applications of vision-based ubiquitous health and activity monitoring system for humans and livestock.

\subsection{Computer Vision in Human Health Monitoring}
\label{subsec::human_cv}

The unobtrusive nature of computer vision instrumentation has made it a favorable sensing modality for monitoring human health and activity across a long period of time. There is a wealth of health insights to be derived from the visual data of how humans move and interact with the environment. One such application is in posture and gait analysis, where vision-based models detect motion abnormalities associated with neurological disorders such as Parkinson’s disease ~\cite{kour2019computer,morinan2022computer, rupprechter2021clinically, archila2022multimodal}, track the progress of stroke rehabilitation ~\cite{thopalli2024advances, rahman2022ai, zhao2022testing}, and assess fall risk in elderly populations ~\cite{yu2017computer, harrou2017vision, feng2014fall, joshi2017fall, gaya2024deep}. Furthermore, computer vision is capable of extracting minute or hyperspectral signals that are impercpetible by mere visual inspections. Vital sign monitoring using remote photoplethysmography (rPPG) and heart beat detection enable contactless measurement of heart rate and respiratory rate by analyzing subtle skin color variations ~\cite{de2013robust,  lokendra2022and, monkaresi.etal2014MachineLearning, arppana.etal2021RealTime, balakrishnan.etal2013DetectingPulse}. This is particularly useful for detecting the onset of respiratory distress conditions like sleep apnea ~\cite{akbarian2020distinguishing, chen2023self}, or pneumonia in children ~\cite{chagas2021new}. Computer vision also plays a role in mental health assessment, where vision-based emotion recognition systems analyze facial expressions, eye movement, and micro-expressions to detect signs of stress, fatigue, or depression ~\cite{singh2021decoding, samareh2018detect, kogilathota2021iponder, kargarandehkordi2023computer, vanneste2021computer}. In fitness and lifestyle tracking, activity recognition systems use pose estimation to monitor daily movement patterns, optimizing exercise routines and detecting sedentary behavior ~\cite{carlson2020automated, mo2016human, khanal2022review, saponaro2019estimating}.

\subsection{Computer Vision in Livestock Monitoring}
\label{subsec::livestock_cv}

In livestock production, one of the most routine operations a population counting of the animals, and tracking their physical growth and well-being. In place of manual inspection methods, computer vision systems have been used for detection, identification, and location tracking 
~\cite{chemme.etal2024ChickenDetection, tassinari2021computer, oliveira2021review, seo.etal2019TouchingPigSeparation,tiwari.etal2021CVDLCattle, kaiyu.etal2023LivestockYOLOV5} by leveraging distinct physical features, such as the muzzle pattern in cattle ~\cite{kaur2022cattle}, or ID markers on plastic eartags ~\cite{alomair.etal2024LivestockOCR}. Other salient physical properties that are indicative of animal growth, and consequently the production yield, are the livestock's body size and weight. Researchers have explored predictive vision frameworks capable of estimating the animals' body dimensions and weight using stereo vision and machine learning techniques ~\cite{ma.etal2024CVLivestockBodyDimensions, dohmen2022computer, bezsonov2021breed, rudenko2020cattle, wang2023review, tscharke2013review, cominotte2020automated}. This insight can help farm operators provide timely interventions if signs of malnutrition or gorging is detected. In addition to statically analyzing the animals features, CV aids in the early detection of livestock health issues by analyzing changes in posture, movement, and feeding behavior to identify signs of illness, such as lameness or respiratory distress ~\cite{evangelista.etal2023DLPoultry, qazi.etal2024AnimalFormer, biglari.etal2022CattleWaterIntake}.

\subsection{Challenges in Health Monitoring via Computer Vision}
\label{subsec::challenges_cv}

In spite of the promising applications of computer vision for health monitoring, several challenges must be addressed to ensure reliable and ethical implementation. One major concern is data quality and variability, as environmental factors such as lighting conditions, occlusions, and background noise can significantly impact image clarity and model accuracy~\cite{oliveira2021review,chemme.etal2024ChickenDetection, tiwari.etal2021CVDLCattle, kaiyu.etal2023LivestockYOLOV5, wang2023review}. In livestock applications, indoor facilities may introduce lighting challenges due to insufficient illumination and shadow occlusions from the animals. Additionally, model robustness and accuracy remain critical barriers, as vision models must generalize across diverse populations, accounting for variations in body size~\cite{ma.etal2024CVLivestockBodyDimensions, tscharke2013review, cominotte2020automated}, posture~\cite{tagoe.etal2024AnimalMovement, qazi.etal2024AnimalFormer}, breed~\cite{bezsonov2021breed, rudenko2020cattle}, and individual health conditions.

\section{Radio Frequency and mmWave}
\label{sec::rf}

Radio frequency (RF) and millimeter wave (mmWave) technology have emerged as promising modalities for pervasive sensing in both livestock and human health monitoring. Most notably, RFID has been widely deployed in agricultural settings to enable systematic tracking and health management of livestock~\cite{ariff.etal2014RFIDLivestockHealth, vrbancic.etal2010RFIDLivestockBreeding, leong.etal2007UHFLivestockIdentification}. Millimeter wave radar techniques, on the other hand, have shown to be well-suited for non-contact human vital sign monitoring, as demonstrated in studies on cardiac activity~\cite{vasu.etal2009DetectionCardiac, iyer.etal2022MmWaveRadarbased}.
Recent advances in deep learning-based radar sensing and wireless signal processing for stress monitoring~\cite{ha.etal2021WiStressContactless} extend the potential of RF systems beyond conventional applications. We aim to leverage these insights to assess and compare the effectiveness, limitations, and integration challenges of RF and mmWave sensing approaches for livestock health and activity monitoring in modern smart farm environments.

\subsection{RF Methods in Human Health Sensing}
\label{subsec::human_rf}

RF-based techniques have paved the way for non-invasive and contactless monitoring of human physiological signals, offering a comfortable alternative to wearable sensors. For example, 5.8 GHz RF sensors are employed to detect subtle cardiac activities, illustrating how electromagnetic signals can capture heart function without direct skin contact~\cite{vasu.etal2009DetectionCardiac}.

In parallel, mmWave radar systems have demonstrated the ability to simultaneously track heart and respiration rates with impressive accuracy, aided by machine learning models that enhance signal processing and arrhythmia detection~\cite{iyer.etal2022MmWaveRadarbased}. Beyond cardiac monitoring, RF methods are shown to have been adapted for stress assessment and other vital sign measurements; contactless seismocardiography, for instance, leverages radar signals to monitor cardiovascular dynamics under varying conditions~\cite{ha.etal2020ContactlessSeismocardiography}, while other studies have extended these techniques to gauge stress levels using wireless signal fluctuations~\cite{ha.etal2021WiStressContactless}.  
 
Further research in this field has explored the use of RF sensing to capture gait patterns, detect falls, and assess overall activity levels, demonstrating its versatility in both clinical and everyday environments. These methodologies provide high-resolution data and also help overcome critical issues related to privacy and user compliance, making them especially valuable for continuous health monitoring in the modern age.

\subsection{RF Methods in Animal Health Sensing}
\label{subsec::animal_rf}

RF methods, particularly those based on RFID technology, have played a key role in modern livestock health management and monitoring. For example, employing RFID to streamline animal identification and real-time health record keeping~\cite{ariff.etal2014RFIDLivestockHealth}. RFID are able to significantly enhance the efficiency of livestock breeding processes by reducing costs and improving animal management. In addition, deployment of both High Frequency and Ultra High Frequency RFID tags highlighted the potential to read multiple tags simultaneously in dynamic farm environments~\cite{leong.etal2007UHFLivestockIdentification}. It has also been examined how RFID-based RF methods can be optimized to achieve robust and reliable livestock health sensing in modern smart farm settings.

\subsection{RF Challenges in Smart Farm Animal Monitoring}
\label{subsec::rf_challenges}

Despite the proven benefits of RF systems in livestock management, several challenges still persist in practical smart farm deployments. The primary issue behind this approach is environmental interference~\cite{leong.etal2007UHFLivestockIdentification, vasu.etal2009DetectionCardiac, iyer.etal2022MmWaveRadarbased, ha.etal2020ContactlessSeismocardiography, ha.etal2021WiStressContactless}. Farms are complex environments with a mix of physical obstacles, variable weather conditions, and electromagnetic noise, all of which can significantly disrupt RF signals. This limits not only read ranges, but also exacerbates tag collisions when numerous animals are present. 

Another major concern behind using RF is scalability. Large-scale smart farm operations often involve thousands of animals, and ensuring consistent, reliable communication among all RFID tags becomes a logistical and technical challenge~\cite{leong.etal2007UHFLivestockIdentification}. Tag collisions where multiple tags attempt to communicate simultaneously can lead to data loss and inaccuracies. Moreover, the robustness of RFID systems can be compromised by the durability of the tags themselves; tags are often exposed to harsh environmental conditions, including moisture, dirt, and physical impacts, which can degrade their performance over time.

Integration with existing farm management systems also introduces further complications. Bridging the gap between raw RF data and actionable insights requires sophisticated data processing and real-time analytics, which are not always readily available in traditional farm setups~\cite{s20092495}. While RF systems are relatively cost-effective, the cumulative expense of deploying and maintaining a large number of sensor nodes and the necessary supporting infrastructure can be significant and in many cases not justifiable. Finally, the continuous evolution of RF technologies means that systems must be adaptable to new standards and protocols, creating ongoing challenges in system upgrades and interoperability~\cite{9422337}. Overall, while RF methods offer considerable promise for livestock health monitoring, addressing these challenges is essential for their successful, long-term application.

\section{Wearables}
\label{sec::wearables}

In addition to the methods described in ~\ref{sec::vib}, ~\ref{sec::cv}, and ~\ref{sec::rf}, wearables have long been used for ubiquitous health monitoring of humans, wild animals, and even livestock. 

\subsection{Wearables in Human Health Monitoring}
\label{subsec::human_wear}

Wearable technology has revolutionized health monitoring in humans by enabling continuous real-time tracking of vital signs and physiological metrics. Devices such as smartwatches, fitness trackers, and medical grade wearables incorporate sensors to measure heart rate~\cite{rukasha.etal2020MedicalWearableHR, zhang.etal2015Troika, anliker.etal2004Amon}, blood oxygen levels~\cite{zhao.etal2021WearableSaturationMonitor, liou.etal2023WearableSp02PR}, body temperature~\cite{thiyagarajan.etal2022MWCNTWearableTemp, yu.etal2020StrainInsensitiveWearableTemp}, and even electrocardiogram (ECG) data. Advanced wearables can detect irregular heart rhythms~\cite{sabor.etal2022WearableECGArrhythmia}, track sleep patterns~\cite{abdulsadig.etal2022AccelerometerNeckSleepPosture}, and monitor stress levels using biometric signals~\cite{mokaya2018myovibe,mokaya2016burnout,mokaya2015acquiring}. In medical applications, portable glucose monitors help diabetic patients manage blood sugar levels~\cite{siddiqui.etal2018WearableGlucoseSurvey}, while smart patches and biofeedback devices assist in the management of chronic disease~\cite{yang.etal2021WearableChronicDiseaseSurvey}. These technologies not only empower individuals to take proactive control of their health but also provide healthcare professionals with remote monitoring capabilities, improving early disease detection, personalized treatment, and overall patient outcomes.

\subsection{Wearables in Tracking Wild Animals}
\label{subsec::wild_wear}

Concerning wild animals, wearable sensors have been used to survey wild zebras~\cite{zhang.etal2004Zebranet}, deer~\cite{jain.etal2009wildCENSE}, dolphins~\cite{shorter.etal2017DolphinBiologging}, birds~\cite{mu.etal2023BirdFlight}, elephants~\cite{fazil.etal2018IoTElephantDetection}and butterflies ~\cite{lee.etal2021MSail} among other species. These systems provide valuable insight to conservationist groups, allowing scientists to track the health on entire populations of wild animals without frequent human intervention. Wearables have also been utilized in anti-poaching efforts, with GPS-enabled collars acting as alarms, letting human conservationist groups know which animals are at risk of being poached in real-time. Given that many animal species frequently change location based on the amount of available resources, nearby predators, and season, utilizing wearables has often been the default methodology for surveying wild animal populations.

\subsection{Wearables in Tracking Livestock}
\label{subsec::livestock_wear}

Wearable devices for tracking livestock offer the benefit of continuous and individualized monitoring of health and behavior and provide real-time data on vital signs such as heart rate and activity levels, which can facilitate early detection of health issues and guide timely interventions~\cite{anand.etal2022RealTimeLivestockRearing, kaur.etal2022IoTMLCattlePrediction}.

Attaching wearables to livestock allows movement patterns and behavior monitoring and helps in assessing overall well-being and optimizing management practices. For instance, the continuous collection of activity data can reveal insights into grazing habits, social interactions, and daily routines, all of which are crucial for maintaining high standards of animal welfare ~\cite{alipio.etal2023GoatFeedingWearable, nagl.etal2003WirelessCattle,liu.etal2024BovineWellBeing,ezhilarasan.etal2023CattleHealthIoT, mekruksavanich.etal2023SensorCattleBehaviorClassification, lavanya.etal2023IoTLivestockMonitor}. 

Furthermore, the granularity of data obtained allows for a more tailored approach. This supports immediate health evaluations and also contributes to long-term strategies in nutrition management and disease prevention ~\cite{alipio.etal2023GoatFeedingWearable, liu.etal2024BovineWellBeing, mekruksavanich.etal2023SensorCattleBehaviorClassification}. 

In summary, wearables in livestock tracking represent a significant step forward in agricultural technology, providing detailed, continuous monitoring that enhances the ability to manage livestock health effectively and efficiently.

\subsection{Challenges with Deploying Wearables in Smart Farms}
\label{subsec::wearable_challenges}

However, monitoring livestock in smart farms via wearable sensors is made difficult by a few factors.

Unlike other sensing methods, wearables scale poorly with livestock occupancy~\cite{alipio.etal2023GoatFeedingWearable, nagl.etal2003WirelessCattle, ezhilarasan.etal2023CattleHealthIoT, mekruksavanich.etal2023SensorCattleBehaviorClassification, lavanya.etal2023IoTLivestockMonitor}. For example, modern smart farms can raise thousands of pigs at one time, with each sow and piglet requiring at least one sensor attached to their body for data collection. Dense sensor deployment can strain the smart farm's network, as the sheer number of sensor nodes all attempting to send data at once can lead to increased packet loss, resulting in inconsistent biometric monitoring~\cite{Polastre2004}. Each sensor node also requires a dedicated battery, meaning that, each time a node depletes its available power, a worker must physically replace the battery on the device in order to restore the stream of biometric data, which is both costly and laborious given the sheer number of animals being raised at once.

Wearables are often struggle in harsh environments due to the sensor node's physical fragility~\cite{zhang.etal2004Zebranet, alipio.etal2023GoatFeedingWearable, nagl.etal2003WirelessCattle, ezhilarasan.etal2023CattleHealthIoT, mekruksavanich.etal2023SensorCattleBehaviorClassification, lavanya.etal2023IoTLivestockMonitor}. Due to discomfort and lack of care, these wearables are often torn off or crushed by the monitored animal, necessitating physical replacement of the sensor node. This issue is only exacerbated by the aforementioned scalability problem, as smart farms must not only maintain the hundreds to thousands of active sensor nodes, but also must stockpile thousands of additional nodes to replace those crushed during deployment. 

As a result of these issues, modern smart farms have largely adopted pervasive sensing methods, such as those described in Sections ~\ref{sec::vib}, ~\ref{sec::cv}, and ~\ref{sec::rf}. Unlike wearables, these modalities are often able to report biometrics for multiple animals per node, and can be deployed out of the animal's reach, resulting in a more efficient, cost-effective, and fault-tolerant sensor deployment. 

\section{Discussion}
\label{sec::disc}

Among all the modalities discussed, each has their unique advantages and disadvantages for deployment in smart-farms. While wearables can work well in when tracking a limited number of animals,modern smart-farms often raise thousands of livestock at once. Since each animal would require its own device, wearables quickly become unfeasible due to the logistics and cost behind maintaining these thousands of devices.

That leaves smart farms with radio-frequency, computer vision, and vibraiton, among other modalities. Much like wearable sensors, RF methods, particularly RFID-based systems, still present a scalability problem, not only becuase of upkeep, but also due to the fact that thousands of bespoke devices would all crowd the same frequency band, potentially leading to data loss, which may be aggravated by other environmental factors. 

While computer vision can monitor many subjects at once, it is plagued by its dependency on environmental lighting conditions, but also the sheer volume of data produced relative to other modalities. A camera recording all day can easily produce tens to hundreds of gigabytes (GB) of data. Since computer-vision based deployments often use several cameras, all recording constantly, researchers have to sift through potentially terabytes of data, whose storage places a financial burden on the research budget. Computer vision is also privacy-invasive, as cameras could capture invasive information concerning the farm's location, its workers, and other senstivite data by accident.

Lastly, vibration has appeared as a viable medium for long-term monitoring in the last twenty years. Unlike wearable devices, RF, and computer vision based methods, vibration sensing is not only scalable, but also preserves the perception of privacy. Relative to the other discussed sensor modalities, it aso produces a relatively small amount of data, on the order of 10-15 GB per day, making storage of said data much more convienent.

However, there remain unsolved problems in utilizing structural vibrations for long-term monitoring. One such problem concerns the need for labeled data, as the lack of comprehensive open-source vibration datasets means that training robust models for vibration sensing is difficult. 

Despite the modalities flaws, we believe that vibration sensing represents the most promising approach for the future of pervasive livestock health and activity monitoring, as while it has unsolved questions that require exploration, the nature of the signal itself avoids many of the issues present when using other sensor modalities.

\section{Conclusion}
This survey has examined a spectrum of sensing modalities with each presenting unique advantages and challenges.
Our review reveals that structural vibration sensing stands out as the most promising approach. 
Vibration-based techniques offer a privacy-preserving, economical, and efficient means of monitoring both activity and vital signs. The inherent advantages of vibration sensing such as its ability to capture subtle biomechanical signals while operating reliably in the complex physical environment of a ``smart'' farm  make it an optimal solution for early detection of health issues and real-time animal management.

Despite challenges in signal propagation and the need for refined machine learning algorithms to handle diverse livestock profiles, the benefits of vibration sensing, including its scalability and minimal intrusion, strongly position it as the modality of choice for future smart farm implementations. Our findings advocate for further research and development in vibration sensing technologies, with an emphasis on integrating advanced signal processing and data fusion techniques to fully exploit its potential in enhancing livestock welfare and operational efficiency.
This survey lays a comprehensive foundation for further interdisciplinary research, encouraging collaboration between engineers, computer scientists, and agricultural experts to pioneer next-generation smart farm solutions.

\balance
\printbibliography

\end{document}